\documentclass[twocolumn,secnumarabic,amssymb, amsmath, nofootinbib,tightenlines,
nobibnotes, aps, prl]{revtex4}
\usepackage{amsmath}%
\usepackage{longtable}%
\usepackage{bm}%
\expandafter\ifx\csname package@font\endcsname\relax\else
 \expandafter\expandafter
 \expandafter\usepackage
 \expandafter\expandafter
 \expandafter{\csname package@font\endcsname}%
\fi

\usepackage{graphicx}
\usepackage{graphicx}

\begin{document}

\title{Skin friction in  zero-pressure-gradient boundary layers.}

\author{Victor Yakhot}
\email{vy@bu.edu}
\affiliation{Department of  Mechanical Engineering, \\
Boston University, Boston, MA 02215}
\date{today}%

\begin{abstract} 
\noindent     A global approach leading to  a self-consistent solution to the Navier-Stokes-Prandtl equations for zero-pressure-gradient boundary layers is presented.   It is shown that as $Re_{\delta}\rightarrow \infty$, the dynamically defined boundary layer thickness $\delta(x)\propto x/\ln^{2}Re_{x}$ and the skin friction  $\lambda=\frac{2\tau_{w}}{\rho U_{0}^{2}}\propto 1/\ln^{2}\delta(x)$.
Here $\tau_{w}$ and $U_{0}$ are the wall shear stress and free stream velocity, respectively. The theory is formulated as an expansion in powers of a small dimensionless parameter $\frac{d\delta(x)}{dx}\rightarrow 0$ in the limit $x\rightarrow \infty$. 
 \end{abstract}
  
  \maketitle

\noindent   The law of  variation of  skin friction with Reynolds number 
in  turbulent wall flows  is one of the oldest riddles  of physics of turbulence.  In addition to difficulties associated with a general problem of strong  isotropic turbulence,
the presence of solid walls  is responsible for  appearance  of two  different characteristic velocities. The so called friction velocity,  reflecting properties of the near-wall sublayer,  is equal to  $u_{*}^{2}=\nu|\frac{\partial U(y)}{\partial y}|_{wall}$ and  the  externally prescribed  free-stream velocity   is denoted  as $U_{0}$.
For  dimensionless distance to the wall $y_{+}=\frac{yu_{*}}{\nu}=O(1)$,   the dimensionless velocity profile  $U_{+}(y_{+})=U(y_{+})/u_{*}\propto y_{+}$ is independent  upon the Reynolds number.  It is the interplay of the two characteristic velocities that makes derivation of mean velocity distributions in wall flows so difficult.
 
\noindent The first semi-empirical theory of zero-pressure-gradient boundary layer (BL) was developed  by Prandtl (for detailed description see Ref.[1]) and references therein)  who, defining self-similar variables $U(x,y)=U(\frac{y}{\delta(x)})$,  introduced the boundary layer thickness $\delta(x)$ depending upon distance to the origin $x$.  In addition, from the force balance, Prandtl  obtained  the expression for skin friction $\lambda\propto \frac{d\delta}{dx}\propto (\frac{u_{*}}{U_{0}})^{2}$.  
Then,  as a stroke of genius, based on  experimental data on skin friction in   {\it pipes and channels} $\lambda_{pipe}\propto Re_{D}^{-\frac{1}{4}}$ available in those early days, he, identifying $\delta(x)$ with the pipe diameter $D$,  proposed  the  differential relation:

$$\frac{d\delta (x)}{dx}\propto (\frac{\nu}{\delta U_{0}})^{\frac{1}{4}}$$

\noindent giving $\delta(x)\propto \frac{x}{Re_{x}^{0.2}}$, and $\lambda\propto Re_{x}^{-0.2}$ \ - \ the expressions  often used  in modern engineering literature and textbooks.   A different, more detailed,   formulation was based on a a force balance following the equations of motion: 

$$\lambda\propto \frac{d\delta(x)}{dx}\int_{0}^{\delta(x)}\frac{U(y)}{U_{0}}(1-\frac{U(y)}{U_{0}})dy$$

\noindent Then substituting into this expression the assumed velocity profile $\frac{U(y)}{U_{0}}\propto \phi(\frac{y}{\delta(x)})$  with either algebraic or logarithmic expressions for the function $\phi(z)$  led to different relations  for skin frictions  reasonably well  representing experimental data in a restricted range of Reynolds number variation.  Due to the lack of mathematically sound derivation, even today, the question of  "correct"  velocity distribution in fully developed pipes/channel  flows is a  topic of   lively scientific discussions.  These  early theories,  based on bold assumptions combined with  deep physical insights, led to helpful expressions widely used in engineering. 

\noindent  
The modern analysis of pipe/channel  and boundary layers is typically based on an   assumed scaling relation for velocity  represented in the "inner" and "outer" regions of the flow as 
$ U(y)=u_{*}f(y_{+})$ and $U(y)=U_{cL}-u_{0}g(\frac{y}{H})$, respectively$^{2}$. 
The parameters $u_{*}$ and $u_{0}$ are corresponding characteristic velocities.   Then, different matching conditions  applied   to  the "overlap"  region lead  to different shapes of velocity profile $U(y)$. According to experimental data$^{2}$,  in the limit $Re_{D}\rightarrow \infty$, the second characteristic velocity $u_{0}\rightarrow u_{*}$.  In a  recent paper$^{3}$   assuming the   logarithmic velocity profile across  a  zero-pressure-gradient boundary layer,   developed an asymptotic expansion,  leading to the so called Coles-Fernholtz relation$^{4}$:
 
 \begin{equation}
 \lambda\propto \frac{1}{\ln^{2} Re_{\delta(x)}}
 \end{equation}

\noindent  widely accepted as an accurate large Reynolds number ( $Re_{D}>>10^{5}$ ) asymptotics. While this work is based on a solid mathematical analysis, its starting point, logarithmic profile, is an assumption not following   the Navier-Stokes equations.

\noindent The theory developed below is based on the following observation. It is well known that in the small-parameter-lacking homogeneous and isotropic turbulence evaluation of moments of velocity fluctuations is an extremely difficult and unsolved problem. In this regard the problem of statistical properties of wall turbulence is at least as difficult. However,
 in wall flows  as $Re_{\delta}\rightarrow  \infty$, the {\it global dimensionless parameters}  $u_{*}/\overline{U}\rightarrow 0$ and $\frac{d\delta(x)}{dx}\rightarrow 0$  are small and can be used for construction of the well-behaved perturbation expansion leading to prediction of  global properties of wall flows.  This is the main goal of this paper.

\noindent    We consider a flat plate $0\leq x\leq \infty$ and $y=0$. The freestream velocity of incoming flow is ${\bf U}_{0}=U_{0}{\bf i}$ and we are to analyze the Navier-Stokes -Prandtl equations in the boundary layer  approximation:
 
 \begin{equation}
 \frac{\partial U}{\partial x}+\frac{\partial V}{\partial y}=0
 \end{equation}

 \begin{equation}
 U\frac{\partial U}{\partial x}+V\frac{\partial U}{\partial y}=\frac{\partial}{\partial y}(\nu\frac{\partial U}{\partial y}+\tau_{x,y})
 \end{equation}

\noindent with $\tau_{x,y}=-\overline{u_{x}u_{y}}$ and the mean is taken over ensemble of fluctuating velocity field ${\bf u}$.  Interested in the global, integral,  features of a  flow we neglect contribution coming from extremely thin in the limit $Re_{\delta}\rightarrow \infty$ viscous sublayer  $0\leq y\leq y_{sL}\rightarrow 0$ and 
as  $x\rightarrow \infty$ we,  assuming  self-similarity of velocity profile write:
$U(x,y)=U(x, \frac{y}{\delta(x)})\equiv U(x, \eta)$, $V=V(x,\frac{y}{\delta(x)})\equiv V(x, \eta)$ and $\tau_{x,y}=\tau_{x,y}(x, \frac{y}{\delta(x)})\equiv \tau_{x,y}(x, \eta)$ where the defined below  width of the boundary layer $\delta(x)$ {\it must be found from equations of motion}.   The incompressibility constraint (2) gives:

\begin{eqnarray}
V(x,y)=-\int_{0}^{y}\frac{\partial U(x,y')}{\partial x}dy'=\nonumber \\
-\delta\int_{0}^{\eta}\frac{\partial U(x,\eta')}{\partial x}d\eta'+\frac{\partial \delta(x)}{\partial x}\int_{0}^{\eta}\eta'\frac{d U(x,\eta')}{d\eta'}d\eta'
  \end{eqnarray}





\noindent Integrating  (3)  over the interval $0\leq y\leq \infty$,  and introducing the `displacement thickness'  $\theta$ we, using (4),  express  the skin friction in terms of the boundary layer thickness  $\delta$:

\begin{eqnarray}
\frac{d\theta}{dx}=\int_{0}^{\infty}G(x,\eta)(\frac{\delta}{U_{0}}\frac{\partial U(x,\eta)}{\partial x}+
\frac{\partial \delta(x)}{\partial x}\frac{U(x,\eta)}{U_{0}})d\eta=\nonumber \\
=\frac{u_{*}^{2}}{U_{0}^{2}}= \lambda/2 \hspace{1in}
\end{eqnarray}


 
\noindent  where $u_{*}^{2}(x)=\nu \frac{\partial U(x,y)}{\partial y}|_{0}$ and $G(x,\eta)=1-\frac{U(x,\eta)}{U_{0}}$.  In this work we are interested in the global, integral, features of wall turbulence, like $\overline{U}$ with the negligible contribution from the viscous sublayer $y/\delta<y_{sL}/\delta\rightarrow 0$ in the limit $Re_{\delta}\rightarrow \infty$. 
 Therefore, in what follows we  seek solution  in the form 
 
\begin{equation}
U(x,y)=u_{*}(x)\varphi(\eta)
\end{equation}

\noindent so that the relation (5) becomes:

\begin{eqnarray}
\delta(\frac{d}{dx}\frac{u_{*}}{U_{0}})\int_{0}^{\infty}G(x,\eta)\varphi(\eta)d\eta +\nonumber \\
\frac{\partial \delta(x)}{\partial x}\frac{u_{*}(x)}{U_{0}}\int_{0}^{\infty}\varphi(\eta)G(x,\eta)d\eta\nonumber \\
=\frac{u_{*}^{2}}{U_{0}^{2}} \equiv i_{1}+i_{2}=\lambda/2
\end{eqnarray}

\noindent It will become clear below that in the limit $Re_{x}=xU_{0}/\nu\rightarrow \infty$  the first term in the right side of (5)  is negligibly small and therefore $\lambda=2(\frac{u_{*}}{U_{0}})^{2}\propto \frac{d\delta}{dx}\rightarrow 0$.   
Based on as yet unknown function $\delta(x)$, we  define an  averaged- over- the -boundary -layer  property $\Psi$: \  
$\overline{\Psi (x)}= \frac{1}{\delta}\int_{0}^{\delta}\Psi(x,y)dy=\int_{0}^{1}\Psi(x,\eta)d\eta$.  
 Since at  the edge of a boundary layer $y=\delta(x)$, the  velocity and kinetic energy are $U=U(x,\delta(x))$ and $K=K(x, \delta(x))$, respectively,  the familiar integral balance equations must be somewhat modified.  For example, integrating the differential energy balance equation:

$$
U\frac{\partial K}{\partial x}+V\frac{\partial K}{\partial y}=-\tau_{xy}\frac{\partial U}{\partial y}-{\cal E}+\frac{\partial}{\partial y}(\nu \frac{\partial K}{\partial y}+Q)
$$

\noindent in the interval $0\leq y\leq \delta(x)$ and recalling that

$$\int_{0}^{\delta}V\frac{\partial K}{\partial y} dy=V(\delta)K(\delta)-\int_{0}^{\delta}K\frac{\partial V}{\partial y} dy$$

\noindent  we, using an incompressibility constraint,  derive:

\begin{eqnarray}
\int_{0}^{\delta}\frac{d}{dx}K(x,y)U(x,y)dy+V(x, \delta)K(\delta)=\nonumber \\-\int_{0}^{\delta}\tau_{xy}\frac{\partial U}{\partial y}dy-\delta \overline{{\cal E}}+Q(\delta)
\end{eqnarray}

\noindent where $Q(\delta(x))=\overline{w(\delta)u_{i}^{2}(\delta)}=O(u_{*}^{3})$.  In the limit $\delta\rightarrow\infty$, $V(x,\delta)\rightarrow 0$ and the relation (8) tends to a familiar energy balance (see for example Ref.[5]).  With 

\begin{eqnarray}
V(x,\delta(x))= -\delta\frac{d\overline{U}(x)}{dx}+\frac{d\delta}{dx}(U(x,\delta)-\overline{U}(x))=\nonumber \\-\frac{d}{dx}(\delta \overline{U}(x))+\frac{d\delta}{dx}U(x,\delta)
\end{eqnarray}

\noindent directly following from (4),   the relation (8) takes a very simple form:

\begin{equation}
\frac{d}{dx}(\delta\overline{KU})-K(x,\delta)\frac{d}{dx}(\delta \overline{U})=\int_{0}^{\delta}\tau_{xy}\frac{\partial U}{\partial y}dy-\delta \overline{{\cal E}}+Q(\delta)
\end{equation}

\noindent The left side of (10)  is the energy balance in the cross-section $x$.  As $Re_{x}\rightarrow \infty$ we can define $K_{0}(x)$ so that  the mean: 

\begin{equation}
\overline{KU}=\frac{1}{\delta}\int_{0}^{\delta(x)}K(x,y)U(x,y)dy=K_{0}(x)\overline{U}(x)
\end{equation}

\noindent  In the bulk of the flow $y\gg y_{sL}\rightarrow 0$ dominating the integral  (11), where   the turbulent kinetic energy $K(x,y)$ is a weak function of $y$, the parameter  $(\overline{K}(x)-K_{0}(x))/\overline{K(x)}\ll 1$.  

\noindent {\it Expansion}. Now, recalling that $u_{*}=u_{*}(x)$ depends on the distance to the origin $x$, we introduce a definition of the  boundary layer thickness:

\begin{eqnarray}
U(\delta(x))-\overline{U}(x)\propto U_{0}-\overline{U}(x)= u_{*}\phi_{u}(\frac{u_{*}}{U_{0}})\nonumber \\
K_{0}(x)= u_{*}^{2} \phi_{K}((\frac{u_{*}}{U_{0}})^{2})
\nonumber \\ 
K(\delta(x))= u_{*}^{2}\phi((\frac{u_{*}}{U_{0}})^{2})
\end{eqnarray}

\noindent As $Re_{\delta}\rightarrow\infty$ the dimensionless parameter $\xi(x)=\frac{u_{*}(x)}{U_{0}}\rightarrow 0$ and taking into account that in this  limit  all functional derivatives 

$$\frac{d^{n}U_{+}}{d\xi^{n}}; \hspace{1cm} \frac{d^{n}\frac{K_{0}}{u^{2}_{*}}}{d\xi^{n}}$$

\noindent are finite (for a discussion see below), we derive:

\begin{eqnarray}
U(\delta(x))-\overline{U}(x)\propto U_{0}-\overline{U}(x)\approx  u_{*}\sum_{n=0}^{\infty}\alpha_{n}(\frac{u_{*}}{U_{0}})^{n}\nonumber \\
K_{0}(x)= u_{*}^{2}\sum_{n=0}^{\infty}b_{n}(\frac{u_{*}}{U_{0}})^{2n}\nonumber \\ 
K(\delta(x))\approx  u_{*}^{2}\sum_{n=0}^{\infty}c_{n}(\frac{u_{*}}{U_{0}})^{2n}
\end{eqnarray}

\noindent where $K_{0}(x)$ is defined as:

\begin{equation}
\overline{KU}=K_{0}(x) \overline{U}(x)=u_{*}^{2}(U_{0}-u_{*}\sum_{i=0}^{\infty}\alpha_{i}(\frac{u_{*}}{U_{0}})^{i})\times\sum_{n=0}^{\infty}b_{n}(\frac{u_{*}}{U_{0}})^{2n}
\end{equation}

\noindent   The relations  (14)  is  a {\it definition}  of expansion coefficients $b_{n}$.  Let us demonstrate that  the anzatz (12)-(13),  combined with the energy balance (10),  leads  the well-known empirical relation $\lambda\propto \frac{1}{\ln^{2} Re_{\delta}}$.   Neglecting for a time being the contribution $i_{1}$ to (7) as small (it will be justified  below) we obtain simple estimates:

\begin{equation}
\frac{d\delta}{dx}u_{*}^{2}U_{0}=O(\frac{u_{*}^{4}}{U_{0}})\ll u_{*}^{3}; \hspace{1cm} \frac{d\delta}{dx}u_{*}^{3}=O(\frac{u_{*}^{5}}{U_{0}^{2}})\ll \frac{u_{*}^{4}}{U_{0}};
\end{equation}

\noindent  It will become clear below that as $x\rightarrow \infty$,
$
\delta u_{*}^{2}\frac{du_{*}}{dx}=O(u_{*}^{5}/U_{0}^{2})$.  Substituting the anzatz (12)-(13) into the energy balance (10) and  taking into account the  estimates  (15), we,  equating  the terms of equal  powers of small parameter   $u_{*}/U_{0}$ obtain in the zeroth order:
$$
\int_{0}^{\delta}\tau_{xy}\frac{\partial U}{\partial y}dy=\delta \overline{{\cal E}}+Q(\delta)=O(u_{*}^{3})
$$

\noindent {\it Locally},    the dissipation rate ${\cal E}$ satisfies a well - known universal scaling relation ${\cal E}(y)=\frac{u_{*}^{4}}{\nu}{\cal E}_{+}(\frac{yu_{*}}{\nu})=\frac{u_{*}^{4}}{\nu}{\cal E}_{+}(y_{+})$ which has  recently been  tested experimentally  in pipe flows$^{6-7}$. Thus:

$$\delta\overline{{\cal E}}=u_{*}^{3}\int_{0}^{\delta}{\cal E}_{+}(y_{+})d\frac{yu_{*}}{\nu}=u_{*}^{3}\int_{0}^{R_{*}}{\cal E}_{+}(y_{+})dy_{+}=O(u_{*}^{3})$$

\noindent justifying the above zero-order estimate, provided in the limit $R_{*}\rightarrow\infty$ the function ${\cal E}_{+}$ decreases with $y_{+}$ rapidly enough so that the convergent integral is independent  
of the upper limit.  The production term is:
\begin{eqnarray}
\int_{0}^\delta \tau_{xy}\frac{\partial U}{\partial y}dy=\nonumber  \\
u_{*}^{2}\int_{0}^{1}\tau_{+}\frac{d U(\eta)}{d\eta}d\eta\approx  -u_{*}^{3}\int_{0}^{1}\tau_{+}\frac{d g(\eta)}{d\eta}d\eta=O(u_{*}^{3})\nonumber \end{eqnarray}

\noindent ($\tau_{+}=\tau_{xy}/u_{*}^{2}$ )  where in the limit $Re_{\delta}\rightarrow \infty$ the integral is dominated by  the outer scaling   $U(y)=U_{0}-u_{*}g(y/\delta)$ (Smits (1999)).  The flux contribution $Q(\delta)=O(u_{*}^{3})$.   

\noindent  According to this estimate, contributions to the right  and left side of (10) are balanced separately. {\it  This fact becomes clear if we notice that each term in the  left side of (10) involves  $x$-derivatives  which are proportional to $\frac{d \delta}{dx}\propto (\frac{u_{*}}{U_{0}})^{2}\rightarrow 0$ while each contribution to the right side is $O(u_{*}^{3})=O(1)$. 
Moreover, since in the interval $y>\delta$, all turbulence characteristics sharply decrease to zero,  all integrals over the $y$- coordinate  in the right side are convergent so that the outcome is independent neither of the upper limit $\delta$  nor $\frac{d\delta}{dx}$.  }
The remaining terms are: 

\begin{eqnarray}
\frac{d\delta}{dx}[u_{*}^{2}(U_{0}-u_{*}\sum_{n=0}^{\infty}\alpha_{n}(\frac{u_{*}}{U_{0}})^{n})\times \sum_{i=0}^{\infty}(b_{i}-c_{i})(\frac{u_{*}}{U_{0}})^{2i}]+\nonumber \\
\delta(x)u_{*}^{2}\sum_{n=0}^{\infty}c_{n}(\frac{u_{*}}{U_{0}})^{2n}\times \frac{d}{dx}(u_{*}\sum_{i}\alpha_{i}(\frac{u_{*}}{U_{0}})^{i})=0
\end{eqnarray}

\noindent We see that the first -order in $u_{*}/U_{0}$ terms are  balanced only if $c_{0}=b_{0}$ 
and in the next order   we, neglecting the  asymptotically small $O((\frac{u_{*}}{U_{0}})^{3})$ contributions,  obtain the  differential equation:

\begin{equation}
U_{0}\delta \frac{d u_{*}^{2}}{dx}=\alpha_{1}u_{*}^{3}\frac{d\delta}{dx}
\end{equation}

\noindent where $\alpha_{1}=2\frac{b_{1}-c_{1}}{c_{0}\alpha_{0}}<0$ is  a constant   which must be obtained from a full local theory.  \\

\noindent  It is easy to see  that  the expression: 

\begin{equation}
\lambda=2(\frac{u_{*}}{U_{0}})^{2}=\frac{\kappa}{\ln^{2} \delta}; \hspace{1cm} \lambda \propto \frac{d\delta}{dx}
\end{equation}

\noindent with $\kappa=8/\alpha_{1}^{2}$ is a solution to (17).  Indeed, integrating  (17)  and dividing the outcome by $U_{0}^{3}$, we obtain:

$$
\frac{\lambda}{2}=\frac{|\alpha_{1}|\sqrt{\kappa}}{4\sqrt{2}}\int \frac{d\lambda}{dx}dx=\nonumber \\
\frac{|\alpha_{1}|\sqrt{\kappa}}{4\sqrt{2}}\lambda
$$

\noindent      {\it This result shows that the anzatz (12)-(13) with $\lambda\propto \frac{1}{ \ln^{2} \delta}$ is a self-consistent solution to the Navier-Stokes -Prandtl equations of motion.}

\noindent   Setting for a time being all proportionality coefficients equal to unity, we introducing   $\delta_{0}=\frac{\nu}{U_{0}}$, $Re_{\delta}=\frac{U_{0}\delta}{\nu}$ and $Re_{x}=\frac{U_{0} x}{\nu}$ solve the  the differential equations  (17) with the result:
$
Re_{\delta}[(\ln Re_{\delta})^{2}-2\ln \frac{Re_{\delta}}{e}]= Re_{x}
$  leading to 

\begin{equation}
\delta(x)\propto \frac{x}{\ln^{2} Re_{x}}
\end{equation}

\noindent  in the limit $Re_{\delta}\rightarrow \infty$.

\noindent  The derivation of   asymptotic expressions (18)-(19)  was based on the relation $\frac{d\delta}{dx}\propto (\frac{u_{*}}{U_{0}})^{2}\propto \lambda$ which is obtained from (5) by neglecting the first contribution to the right  side.  With the scaling (6) this is equivalent to omitting 
$i_{1}$  in (7) as small. 
Now, based on the derived solution (18)-(19)  we justify this approximation. Indeed,  comparing the integrands in the right side of (5) gives 

\begin{equation}
i_{1}=\delta(\frac{d}{dx}\frac{u_{*}}{U_{0}})G=\frac{x}{\ln^{2}Re_{x}}\frac{d\sqrt{\lambda}}{dx}G\approx \frac{G}{\ln^{4} Re_{x}}
\end{equation}
 
\noindent and 

\begin{equation}
i_{2}=\frac{d\delta}{dx}\frac{u_{*}}{U_{0}}G\approx  \frac{G}{\ln^{3}Re_{x}}
\end{equation}
 
\noindent so that as $Re_{x}\rightarrow \infty$ the ratio $\frac{i_{1}}{i_{2}}\rightarrow 0$. Thus the term $i_{1}$ in (7) is asymptotically negligible.



\noindent {\it Summary and discussion.} 
 
 \noindent 1. \  In this work the expression for skin friction in the zero-pressure-gradient boundary layer has ben derived directly from the Navier-Stokes equations. Similar results were recently presented in a resent paper$^{3}$.  In this work,  to fit   experimental  data, 
 the authors used a particular   forth-order Pade approximant   yielding  logarithmic velocity profile in   a  leading order.  This immediately gave   $U_{+}(\delta)\approx \ln \delta_{+}$ and  the Coles-Fernholtz  relation (1) for the skin friction. Since it has not been shown that  the Pade approximant is an asymptotic  solution to the Navier-Stokes equations, this theory is essentially semi-empirical.
 
 \noindent 2. \ The theory presented here is different. No experimental information has been used to derive 
 (18)-(19). Moreover, while   we can say  that these expressions are consistent with $U_{+}\propto \ln y_{+}$,  the mathematical proof that this is so does not exist.  
 
 \noindent 3. \ In arriving in  the final result (18)-(19) we assumed that all scaling functions $\phi_{i}(\xi)$ are analytic. It is important to understand that in a  logarithmic theory one can expect singularities when dealing with {\it derivatives over spatial coordinates $x$ or $y$}. It is not what has been done deriving the   expansions  (12)-(13). It is clear that   the derivatives over the {\it functions}  $\xi(x)=u_{*}(x)/U_{0}$ cannot be singular by definition: it is impossible to obtain   a  large response  of, say,  $U_{+}$ upon infinitesimal variation of the friction velocity $u_{*}$.  This simply contradicts all conservation laws.

 \noindent 4. \  The expression (18)-(19)  for friction factor (skin friction)  was found as {\it a self-consistent solution to the Navier-Stokes-Prandtl  equations}. The full mathematical theory leading,  in addition to the scaling relations derived here, to the amplitudes,  remains an unsolved problem.
  
 \noindent 5.\ These results are accurate up to the $O(u_{*}/U_{0})$-corrections.  \\

 \noindent I am grateful to  I. Staroselsky, S.  Bailey,  G. Falkovich, J. Schumacher,  K.R. Sreenivasan,  A. Smits,  A. Polyakov, P. Monkewitz , N. Peters, C. Vassilicos, A. Yakhot  and the students  of the graduate turbulence course (BU, ME 709) for many informative and stimulating discussions.\\

\noindent $^{1}$H. Schlichting, {\em Boundary-Layer Theory}, New York, NY, McGrow-Hill Book Company (1968). \\
$^{2}$A.J. Smits,  \&  I. Marusic,  {\em High Reynolds number flows: a challenge for experiment and simulation}, AIAA 99-3530 (1999).\\
$^{3}$P.A. Monkewitz,  K.A.  Chauhan, \&  H.M.  Nagib  {\em Self-consistent high-Reynolds number asymptotics for zero-pressure-gradient   boundary layers}, Phys. Fluids {\bf 19}, 115101 (2007) .\\
$^{4}$H.H. Fernholtz, \&  P.J. Finley,   {\em The incompressible zero-pressure gradient turbulent boundary layer}, Prog. Aerosp. Sci. {\bf 32}, 245-211 (1966).\\
$^{5}$J.O. Hinze,  {\em Turbulence}, New York, NY, McGraw Hill Book Co. 1965.\\
$^{6}$X. Wu \&  P. Moin,  
{\em A direct numerical simulation study on the mean velocity characteristics in turbulent pipe flow}, 
J. Fluid Mech. {\bf 608}, 81--112 (2008).\\
$^{7}$V. Yakhot, S.C.C. Bailey  \&   A.J. Smits, A.J.,  {\em Scaling of global properties of turbulence and skin friction in pipe and channel  flows}
, J.Fluid Mech., {\bf 652}, 65-73. (2010)\\

\end{document}